**A 20-Channel Magnetoencephalography System Based on Optically Pumped Magnetometers**

Amir Borna[1], Tony R. Carter[1], Anthony P. Colombo[1], Yuan-Yu Jau[1], Christopher Berry[2], Jim McKay[3], Julia Stephen[4], Michael Weisend[5,6], Peter D. D. Schwindt[1]

[1] Sandia National Laboratories, PO Box 5800, Albuquerque, NM 87185-1082, USA
[2] Currently with the Infinera Corporation,140 Caspian Ct., Sunnyvale, CA 94089
[3] Candoo Systems Inc., 2991 Thacker Avenue, Coquitlam BC Canada V3C 4N6
[4] The Mind Research Network and Lovelace Biomedical and Environmental Research Institute, Albuquerque, NM 87106
[5] Rio Grande Neurosciences, Inc., 6401 Richards Avenue, Santa Fe, NM 87508
[6] Department of Neurosurgery, University of New Mexico, Albuquerque, NM 87131

**Abstract**

We describe a multichannel magnetoencephalography (MEG) system that uses optically pumped magnetometers (OPMs) to sense the magnetic fields of the human brain. The system consists of an array of 20 OPM channels conforming to the human subject's head, a person-sized magnetic shield containing the array and the human subject, a laser system to drive the OPM array, and various control and data acquisitions systems. We conducted two MEG experiments: auditory evoked magnetic field (AEF) and somatosensory evoked magnetic field (SEF), on three healthy male subjects, using both our OPM array and a 306-channel Elekta-Neuromag superconducting quantum interference device (SQUID) MEG system. The described OPM array measures the tangential components of the magnetic field as opposed to the radial component measured by all SQUID-based MEG systems. Herein we compare the results of the OPM- and SQUID-based MEG systems on the auditory and somatosensory data recorded in the same individuals on both systems.

**I.      Introduction**

Magnetoencephalography (MEG) measures the magnetic field produced by neuronal currents in the human brain [1, 2]. The most widely used sensor for MEG is the SQUID magnetometer. With this mature commercial technology, arrays of a few hundred sensors are constructed to surround the whole head capturing the signals from cerebral cortex and other brain structures. SQUID-based MEG systems are essential tools for clinical and experimental neuroscience when large scale sensor arrays, millisecond time resolution, and accurate localization of sources within the brain are desired. The importance of SQUID-based MEG systems to neuroscience motivates research into addressing the limitations of these systems. One major limitation is the need for cryogenic liquid helium (He) to operate SQUID-based systems. The Dewar containing the SQUID sensors and the liquid He is formed into a helmet shape to distribute sensors around the head. The Dewar walls of the MEG helmet are ~2-cm thick to provide sufficient thermal insulation. In addition, the helmet is rigid and sized for large adult heads to accommodate the largest number of subjects. Therefore, MEG measurements in individuals with small head size, particularly children, can have many centimeters of head-to-sensor separation. Because dipolar fields decay with the inverse cube of distance, large distances between the sensor array and brain negate the advantage of MEG over electroencephalography (EEG) in source localization [3]. Two recent simulation studies also demonstrate the distinct advantages of moving the sensors closer to the brain [4, 5]. The brain-to-sensor distance and size of the MEG system can be substantially reduced if the need for liquid He is eliminated. Additionally, the size of SQUID-based MEG systems typically requires the use of large, expensive, magnetically shielded rooms (MSR). Eliminating liquid He, and hence the Dewar, could also make the MEG system significantly smaller such that an MSR is unnecessary.

Optically pumped magnetometers (OPMs) are a potential replacement for low temperature (low-$T_C$) SQUID sensors that require liquid He in MEG. In OPMs an atomic gas, typically contained in a glass cell, is illuminated with light that is resonant with electronic transitions in the atom. OPMs require no liquid He and operate at or above room temperature. MEG with an OPM system was first demonstrated by the Romalis group [6, 7]. Their OPM design used large-diameter free-space laser beams to interrogate the atomic sample and a small, person-sized magnetic shield. More recent OPM development for MEG has focused on modular designs [8-11] where light is brought to the OPM either via fiber optics [12-15] or by incorporating a laser into the OPM module allowing flexible placement of the OPM [16]. The small modular OPMs can be constructed in form factors that allow direct contact with the scalp. Highly miniaturized OPMs demonstrate a sensor-to-head distance as small as 4 mm [14]. Another notable sensor being developed for MEG is the high-TC SQUID sensor [17], which operates at liquid nitrogen temperatures, demonstrating sensor-to-head distances of 3 mm [18]. A promising OPM array in demonstrated in [19], and source magnetic localization relative to the brain's anatomy has been accomplished by scanning a single OPM over the scalp [20]. The small modular OPM arrays could be a significant advance for the MEG field by increasing sensitivity to neurological signals. Our group has developed OPM modules for MEG with four spatially separated channels. Recently, we redesigned the sensor to increase the spacing between the four channels to 18 mm and bring the sensing volume closer to the head [21]. In this paper, we report the development of an MEG system, where five modules, forming a 20 channel array partially cover the left side of the head, are placed in a person-sized shield and evoked responses from the auditory and somatosensory cortices are measured.

## II.  Materials and Methods
### a. The Sensor: Optically Pumped Magnetometer (OPM)

The OPM sensors that make up the array are custom built by our group and are described in detail in Reference [21]. We briefly describe the sensors here. In our OPM, a vapor of rubidium atoms is contained within a glass cell, and laser light passes through the cell to optically pump the atoms into a magnetically sensitive state and to probe the atoms' response to an external magnetic field. The sensors use a two-color pump/probe scheme, where the pump and probe laser beams travel collinearly through the sensor. The scheme is an extension of an OPM using elliptically polarized light [12], and operates in the so-called spin-exchange relaxation-free (SERF) regime [22]. In the SERF regime, the OPM is operated at a high density of the rubidium vapor and near zero magnetic field—under these conditions the sensitivity is greatly enhanced to the ~fT/rt-Hz level necessary for MEG.

The OPM sensor's schematic is shown in figure 1(a) this sensor has a footprint on the head of 40 mm x 40 mm, and the distance between the head and the center of sensing volume is 12 mm. The head to sensor distance was increased from the 9 mm described in [21] to add an additional layer of insulation for the comfort of the subject. A polarization-maintaining (PM) fiber delivers the probe (780 nm) and pump (795 nm) laser light to the sensor module. A custom waveplate makes the polarization of the pump laser circular while maintaining the linear polarization of the probe. The pump and probe lasers are split into four beams by a diffractive optical element (DOE), and then the beams pass through a large collimating lens forming 2.5-mm diameter beams directed toward the vapor cell. The optical path length through the Rb vapor is 4 mm. In an individual channel, the volume over which the magnetic field is measured is the intersection of the beam with Rb vapor, ~20 mm$^3$ (Figure 1(b)). The vapor cell's back wall has a high-reflectivity coating to direct the light back to the detection optics. On the return path, the pump laser (795 nm) is blocked by an interference filter and the polarization of the probe light (780 nm) is analyzed by passing the beams through a polarizing beam splitter and subtracting the photodiode outputs of the detected light.

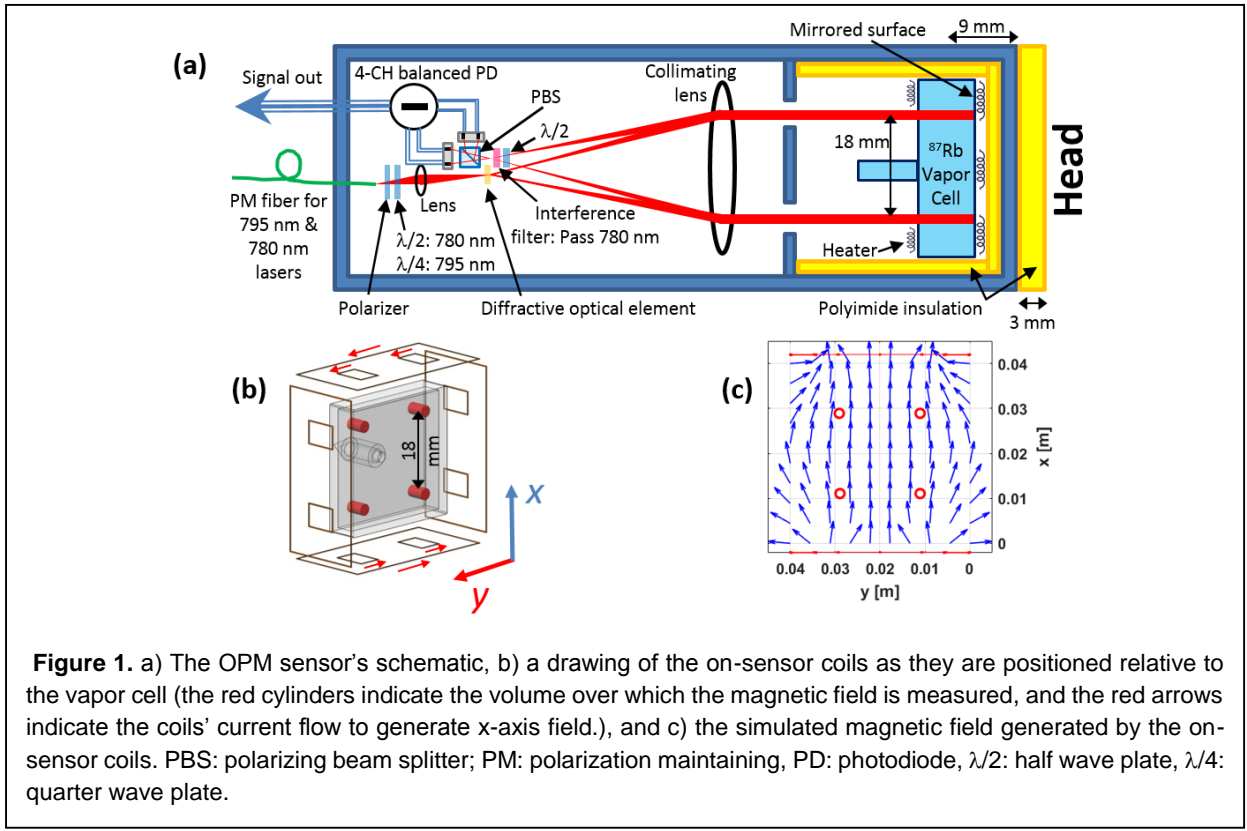

**Figure 1.** a) The OPM sensor's schematic, b) a drawing of the on-sensor coils as they are positioned relative to the vapor cell (the red cylinders indicate the volume over which the magnetic field is measured, and the red arrows indicate the coils' current flow to generate x-axis field.), and c) the simulated magnetic field generated by the on-sensor coils. PBS: polarizing beam splitter; PM: polarization maintaining, PD: photodiode, λ/2: half wave plate, λ/4: quarter wave plate.

The sensitive axis of the magnetometer is defined by application of a modulated magnetic field of amplitude 250 nT and 1 kHz frequency perpendicular to the optical axis of the magnetometer. Lock-in amplifications is then used to demodulate the magnetometer signal to determine the magnetic field. Figure 1(b) depicts the on-sensor modulating coils used to apply this magnetic field and the current flow employed in one pair of coils to measure the field component in the x direction. The other pair of coils is used to apply fields in the y direction. Figure 1(c) shows the simulated magnetic fields of the modulating on-sensor coils; the channel locations are also included in this figure.

### b. The Human-Sized Shield

Due to the large size of the Dewar in SQUID-based MEG devices, these systems typically require a MSR [23]. In comparison, OPM systems can be magnetically shielded using a smaller, and less expensive, multi-layer human-sized shield. Our shield is constructed from three layers of a high magnetic permeability nickel-iron alloy (figure 2). In a cylindrical magnetic shield, fields applied along the longitudinal axis are much less shielded compared to fields applied in the transverse directions [24]. In our design, the longitudinal shielding is further reduced by the need to have an opening for the human subject. For safety and the comfort of the human subject, the subject is never fully enclosed within the shield. We performed finite element analysis of the shield to study geometries that maximize the longitudinal shielding factor (ratio of the external field to the internal field). The diameter of the opening for the human subject is the same size as that of the standard magnetic resonance imaging (MRI) machines, 60 cm, and by adding cylindrical extensions onto these openings, the shielding is increased by a factor of 1.7. More importantly, the first and second order gradients in the longitudinal field are reduced by a factor ~10 in the region where the array is placed. The simulation also shows that if a fourth layer is added while maintaining the dimensions of the internal and external shield layers, the shielding factor is improved by a factor of 6 relative to our current design. The simulated longitudinal shielding factor of our shield at low frequency is 17,000. However, when experimentally measured, the shielding factor at ~0.1 Hz is 1,300. While we have not determined the reasons for this discrepancy, it may be due to imperfections in the geometry of the shield (concave ends where the human subject enters) and reduced permeability of the shielding material from the ideal. The shield has degaussing wires running along the longitudinal direction such that each shield layer is enclosed in a five turn coil. To improve the shielding factor and reduce the residual magnetic field inside the shield, the shield is degaussed before conducting the magnetoencephalography experiments by running a ~20-A, 60-Hz current in series-connected degaussing wires. After degaussing, the residual magnetic field is ~1 nT.

### c. The MEG System Overview

The MEG system's components are shown in figure 2. The laser system provides the probe and pump lasers required to operate the OPM sensors; it is composed of three distributed feedback laser diode sources (Eagleyard, Germany): two pump lasers whose frequencies are separated by 20 GHz and tuned symmetrically about the rubidium D1 line, and a probe laser which is tuned 133 GHz from the D2 line. The pump lasers are combined and amplified to a power level of 1.5 W using a commercial tapered amplifier system (Toptica, Germany). The probe laser also passes through a tapered amplifier to reach a power level of 0.7 W. The amplified pump (795 nm) and probe (780 nm) lasers are combined using a dichroic combiner. The linearly polarized, collinear beams are distributed to three fiber coupling ports with polarizing beam splitters and half waveplates. Each of the fibers delivers light to a polarization maintaining fiber splitter to further divide the light among three fibers. One of the nine fibers is used to monitor and control the probe and pump laser power through adjusting the tapered amplifier's current by two servo controllers.

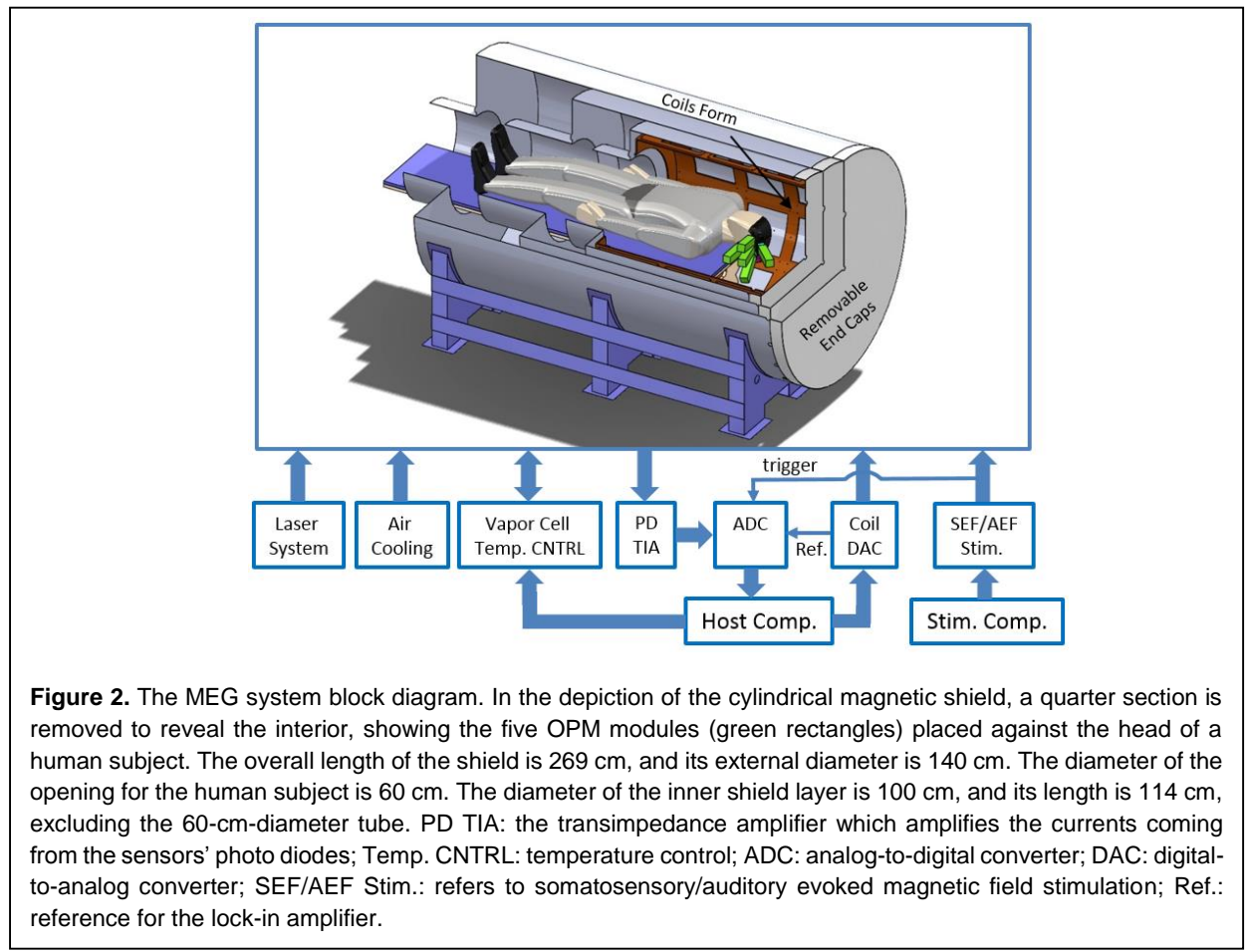

**Figure 2.** The MEG system block diagram. In the depiction of the cylindrical magnetic shield, a quarter section is removed to reveal the interior, showing the five OPM modules (green rectangles) placed against the head of a human subject. The overall length of the shield is 269 cm, and its external diameter is 140 cm. The diameter of the opening for the human subject is 60 cm. The diameter of the inner shield layer is 100 cm, and its length is 114 cm, excluding the 60-cm-diameter tube. PD TIA: the transimpedance amplifier which amplifies the currents coming from the sensors' photo diodes; Temp. CNTRL: temperature control; ADC: analog-to-digital converter; DAC: digital-to-analog converter; SEF/AEF Stim.: refers to somatosensory/auditory evoked magnetic field stimulation; Ref.: reference for the lock-in amplifier.

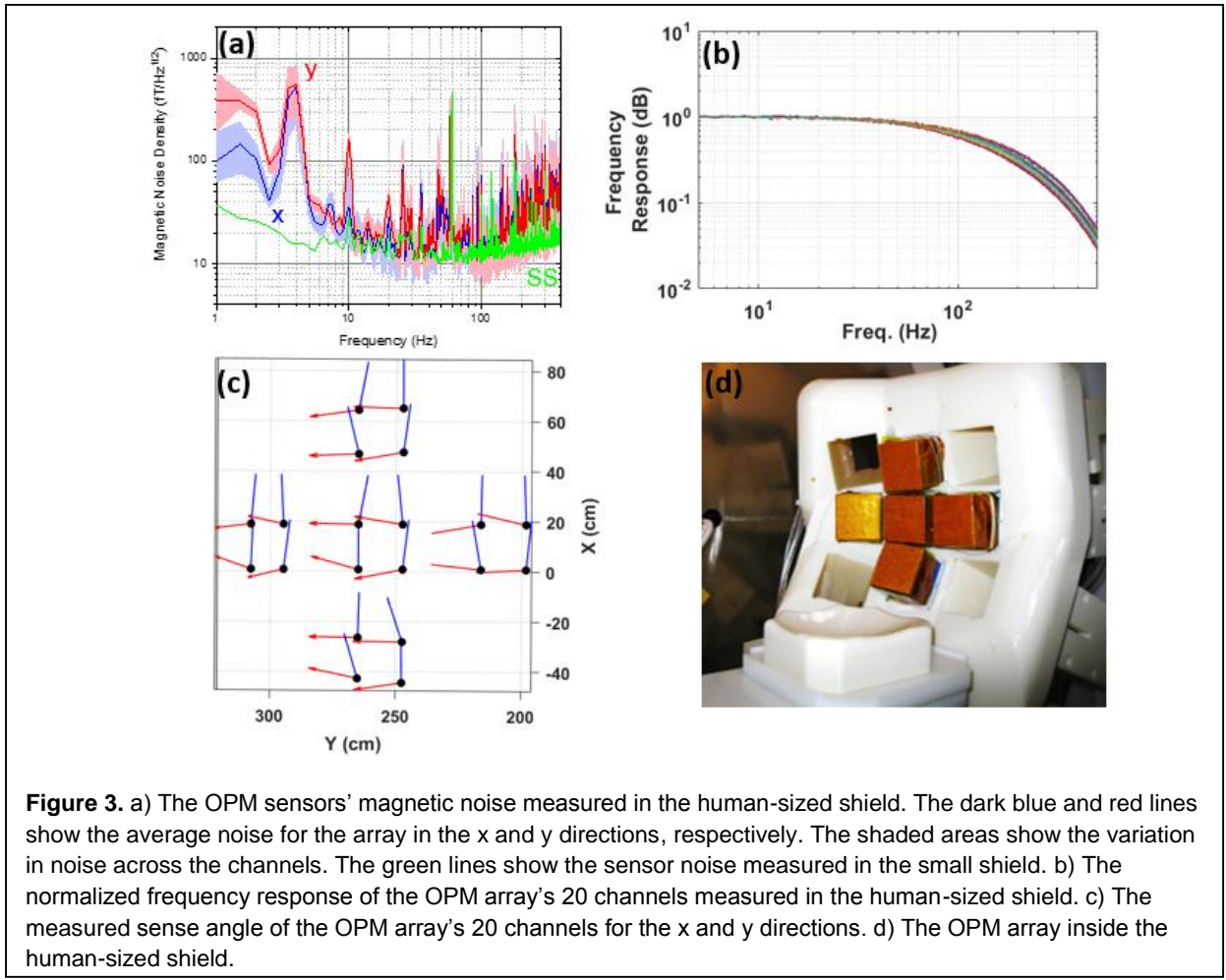

**Figure 3.** a) The OPM sensors' magnetic noise measured in the human-sized shield. The dark blue and red lines show the average noise for the array in the x and y directions, respectively. The shaded areas show the variation in noise across the channels. The green lines show the sensor noise measured in the small shield. b) The normalized frequency response of the OPM array's 20 channels measured in the human-sized shield. c) The measured sense angle of the OPM array's 20 channels for the x and y directions. d) The OPM array inside the human-sized shield.

The temperature controller block independently sets the temperature of the vapor cells with sub 1°C precision using a custom designed servo controller. The sensors' cells are heated up to temperature ranging from 150 to 180°C; hence despite the use of thermal insulation, the surface of the sensors, which will be in contact with the subject's head, can reach to temperatures as high as 45°C. Therefore, air is blown through narrow channels embedded into the 3D printed sensor walls to cool down the sensors' surfaces to room temperature.

The system requires substantial magnetic field control. The coil form (brown cylindrical structure in Figure 2) supports 18 coils to control the magnetic field and first- and second-order gradients within the magnetic shield. With 20 on-sensor coils (four for each sensor), there are 38 coils that must be controlled. A custom coil driver board provides current to the coils, which is in turn controlled by a commercial digital-to-analog converter with 13-bits of resolution and 20 kS/s conversion rate. The coil driver board also provides the reference for the lock-in amplifier. The transfer function of the 18 field coils was measured using a commercial fluxgate; assuming a linear transfer function, the field coils are employed for zeroing the magnetic field and to apply calibrated fields to each sensor to measure the sensors' gains.

All photo diode signals, i.e. 20-channels, are passed to transimpedance amplifiers and digitized using a commercial 48-channel analog-to-digital (ADC) converter with 100 kS/s conversion rate and 24-bit resolution. Using a software-based lock-in amplifier [25] the digitized magnetometer channels are demodulated, filtered with a fourth-order low-pass filter with a time constant of 0.3 ms, averaged and down-sampled to 1 kS/s. The sensed magnetic flux densities are calculated based on the measured gain and stored on the host computer.

Figure 3 shows the OPM array's performance metrics measured inside the human-sized shield. The human-sized magnetic shield is in a laboratory where there is an abundance of magnetic noise from power lines, various instruments, and a nearby freight elevator. Figure 3(a) shows the measured noise on all 20 channels for both x- and y-axes. For comparison, this figure also includes the noise performance of an OPM sensor measured inside a 4-layer, small shield (SS), shown in green, which is superior to that of the array. With the sensors installed in the shield, noise spurs are observed that we primarily associate with building vibrations [26]. To reduce the impact of building vibrations on noise performance, we mechanically isolate the shield from the laboratory's floor using 2-in-thick polychloroprene synthetic rubber (DuPont, US); the added mechanical isolation reduces the average noise by 34 % between 10 and 80 Hz and by 84 % between 1 and 500 Hz. Figure 3(b) shows the measured normalized frequency response of all 20-channels for both x- and y-axes; the sensors' cells are heated up to a predetermined temperature which yields the desired overall bandwidth, i.e. 85 Hz, given the available pump-laser power for each sensor and the low-pass filter of the lock-in amplifier. The channels' 3dB bandwidth vary from 78 Hz to 95 Hz.

For accurate measurement of the magnetic field, we must determine the precise component of the magnetic field being measured or the "sense angle". This is primarily determined by the direction of the 1-kHz modulation supplied by the on-sensor coils. The magnetic field simulation of the on-sensor coils (figure 1(c)) reveals that the sense angles of the sensor's channels are not parallel. With all of the on-sensor coils for a particular direction applied simultaneously and in phase, the field from the neighboring sensors' coils will add to the field from the local on-sensor coils. An additional effect is that the pump laser produces a AC Stark effect that manifests as a fictitious magnetic field parallel to the pump laser propagation direction [27]. While this fictitious magnetic field is largely canceled by appropriate detuning of the two pump laser frequencies [21], imperfect cancellation results in a rotation of the sense angle. To calibrate these effects, we measure the sense angle of each channel by applying a 40 Hz magnetic field and rotating it about the optical axis of a particular sensor using the shield coils. The results of measurements are shown in figure 3(c).

### d. Magnetoencephalography Setup

For MEG measurement using the OPM array, the subject lies in the supine position inside the human-sized magnetic shield; in this position, as shown in figure 2, the OPM array (figure 3(d)) is located next to the left hemisphere of the subject's head covering parts of the auditory, somatosensory, and motor cortexes. The OPM array senses the two field components, i.e. x- and y-axes, in two sequential recording experiments. The protocols of the MEG experiments were approved by the Human Studies Board of Sandia National Laboratories and Chesapeake IRB. We conducted two MEG experiments, auditory and somatosensory, on three healthy male subjects aged between 37 and 42 years old. To have a comparison baseline both auditory and somatosensory MEG experiments were also repeated using a 306-channel Elekta-Neuromag SQUID system (Elekta, Sweden) located in a MSR at the Mind Research Network (Albuquerque, NM).

For the somatosensory experiments a commercial constant current high voltage peripheral stimulator, DS7A (Digitimer, United Kingdom), was used to stimulate the subject's median nerve on the right wrist through two 8-mm felt pads spaced 25 mm apart. Before the experiment, for each subject, the threshold amplitude of the unipolar 200 µs stimulation pulse is determined according to the 2-cm thumb's twitch response. The peripheral stimulator is controlled by a commercial stimulus delivery and experiment control program, Presentation (Neurobehavioral Systems, US), running on the stimulation computer. The peripheral stimulator's trigger is digitized by the same ADC used to collect the MEG data (figure 2) to synchronize the recorded MEG channels with the stimulus timing. The stimulus delivery script sends a total of 400 trigger pulses, with random intervals varying between 1.011 s to 1.04 s, to the Digitimer stimulator. For the auditory MEG experiment the oddball paradigm was employed, in which the test subject is presented with a series of standard tones, 1 kHz, and rare tones, 1.2 kHz, with random intervals varying between 1.04 and 1.54 and a pulse width of 100 ms. The audio tones are presented to the subject using non-magnetic, 50 Ω, Insert Earphones (Etymotic Research, Inc., US). The earphones are controlled directly by the stimulus delivery program which also sends the two triggers, associated with standard/rare tones, to the system's ADC. The OPM array continuously records the MEG signals on all 20 channels. The stimulus delivery script sends a total of 400 (150) trigger pulses for the standard (rare) tones. For both the somatosensory and auditory experiments when measuring with the OPM array, each experiment was run twice, once for each field component being measured, while for the SQUID system, it was run only once.

### III. Magnetoencephalography Signal Processing

For signal processing of the MEG data recorded by the OPM array, the FieldTrip software [28] is employed. The recorded MEG channels are converted to the sensed magnetic flux density using the measured gains for each channel. The continuous MEG channels are then band pass filtered by a two-pass, 4th order Butterworth filter with a high- and low-pass corner frequencies of 0.5 Hz and 150 Hz respectively; this frequency band covers the components of interest for both auditory and somatosensory evoked magnetic fields. The bandpass filter removes the slow varying drifts and DC offsets of the channels. Using the FieldTrip software, a section of channel data without discontinuities is selected, and the rest are discarded. Discontinuities in the MEG data are caused by the subject's movements and if present, can lead to incorrect source separation. Independent component analysis (ICA) [29, 30] is run on the selected continuous block of data, and the noise components, e.g. heartbeat, are removed. In some experiments, i.e. SEF, more ICA components might be removed but the noise and Magnetocardiogram (MCG) are always removed. The continuous MEG channels are then reconstructed from the remaining ICA components, and using the digitized trigger signal, the channels are time-locked and averaged over the remaining trials. Epochs of 1000 ms total duration, with a 300 ms prestimulus period, were time-locked relative to the stimulus triggers. The length of the continuous block of MEG data varies among the three subjects depending on the discontinuity intervals. However, for all the AEF and SEF experiments among the three subjects, the number of averaged epochs was larger than 100.

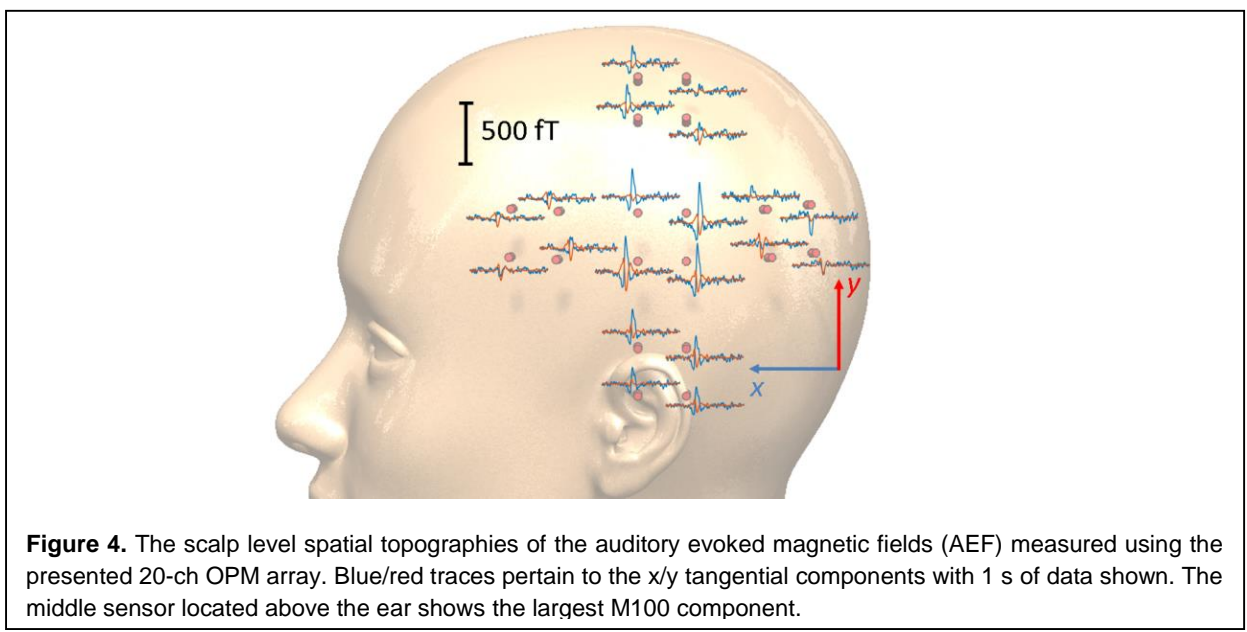

**Figure 4.** The scalp level spatial topographies of the auditory evoked magnetic fields (AEF) measured using the presented 20-ch OPM array. Blue/red traces pertain to the x/y tangential components with 1 s of data shown. The middle sensor located above the ear shows the largest M100 component.

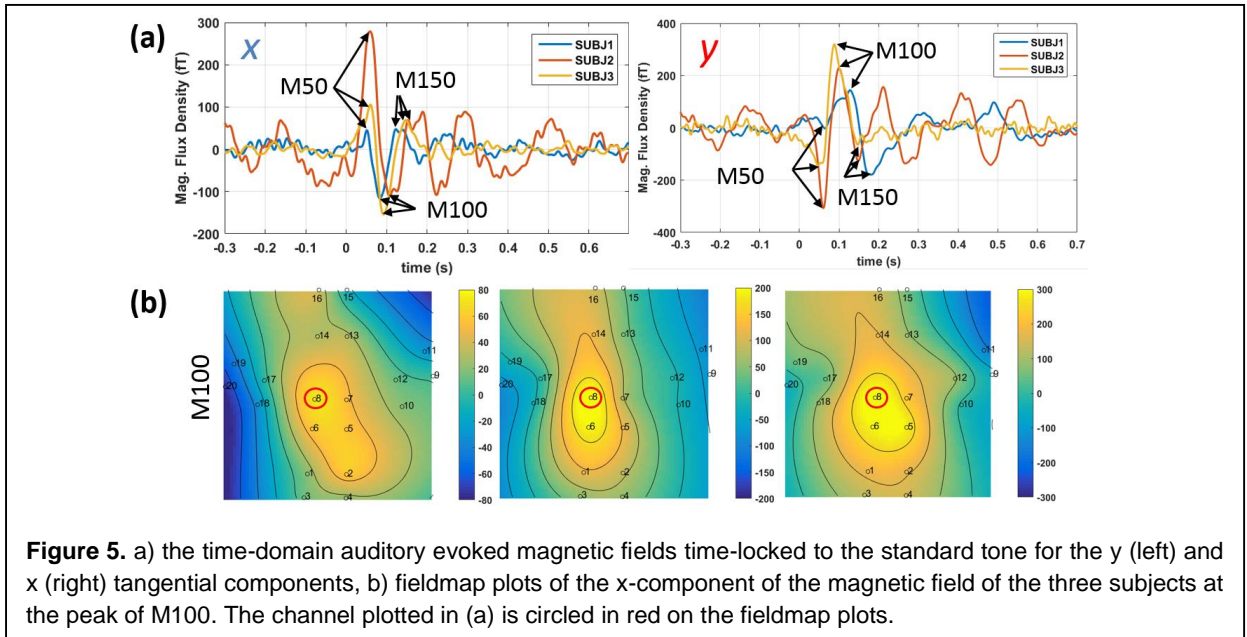

**Figure 5.** a) the time-domain auditory evoked magnetic fields time-locked to the standard tone for the y (left) and x (right) tangential components, b) fieldmap plots of the x-component of the magnetic field of the three subjects at the peak of M100. The channel plotted in (a) is circled in red on the fieldmap plots.

## IV. Results
### a. Auditory Evoked Magnetic Field (AEF)

The averaged response to the standard, 1 kHz, tone for one of the subjects is depicted in figure 4. The AEF signals for both x- and y-axes are strongest for the sensor located over the auditory cortex, i.e. the middle sensor. Because the OPM array measures the components of the field tangential to the head, the magnetic field from a current dipole is maximized directly over the dipole, and we observe maximum roughly where expect.

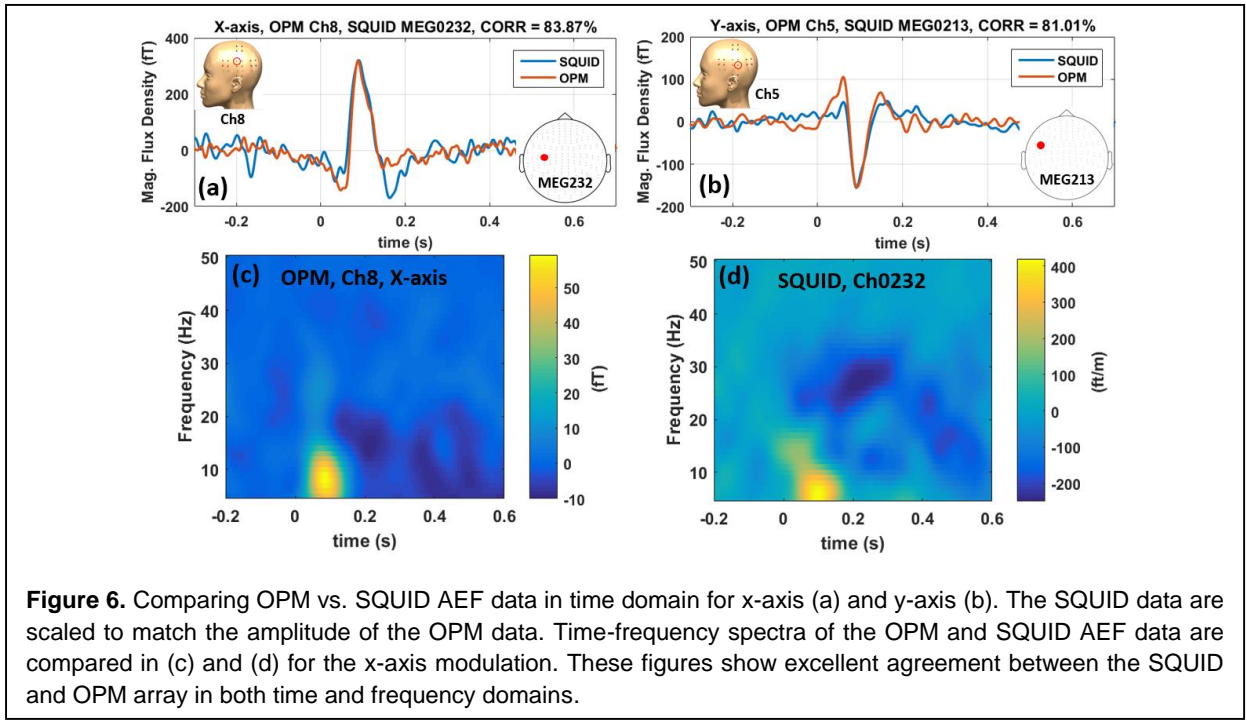

**Figure 6.** Comparing OPM vs. SQUID AEF data in time domain for x-axis (a) and y-axis (b). The SQUID data are scaled to match the amplitude of the OPM data. Time-frequency spectra of the OPM and SQUID AEF data are compared in (c) and (d) for the x-axis modulation. These figures show excellent agreement between the SQUID and OPM array in both time and frequency domains.

Figure 5(a) shows the standard-tone averaged waveform of three subjects for both x (right) and y (left); the middle latency auditory evoked field (MLAEF), M50, can be seen for all the three subjects around 50 ms after the stimulus onset. As expected, the amplitude of the MLAEF response, correlates with the age of the participants [31]. The auditory long latency response field (ALR), M100 (N1m) [32], and M150 are also indicated on both x and y graphs around 100 ms and 150 ms respectively. Figure 5(b) shows fieldmap plots of the x-component of the AEF for the three subjects at the M100 (N1m) peak. We choose the x-component because it has the largest amplitude over the auditory cortex. Based on the placement of the sensor array shown in Fig. 4, the maxima in Fig. 5b are approximately located over the auditory cortex, for all the three subjects.

The comparison between the AEF results of the OPM and SQUID systems is shown in figure 6. In figures 6(a) and 6(b), the standard-tone averaged waveform from one channel with the largest amplitude response, located over auditory cortex, is compared to the SQUID's gradiometer channel located roughly around the same region and with the same orientation. Before averaging, the continuous SQUID data is filtered using a two-pass, 4th order Butterworth filter with a low- and high-pass corner frequencies of 150 Hz and 0.5 Hz respectively; furthermore, notch filters at 60 Hz and 120 Hz are applied to remove the power line noise from the SQUID data. However, unlike the OPM array, the SQUID channels do not undergo independent component analysis, and the SQUID channels are time-locked after filtering. The correlation between the two system's time-domain waveforms are 84% and 81% for x and y orientations respectively. It should be noted that the OPM's channel is a magnetometer whereas the SQUID channel is a planar gradiometer; hence, in this figure, the gradiometer channel is scaled arbitrarily to match the OPM channel. We also compare gradiometer channels that measure the gradient in the same direction as the field component. Interestingly, for a current dipole within a spherical conductor, a tangential field component and a planar gradiometer of the same orientation have qualitatively similar field maps. Figures 6(c) and 6(d) compare the time-frequency spectra of the OPM and SQUID systems showing both systems measure similar spectral content.

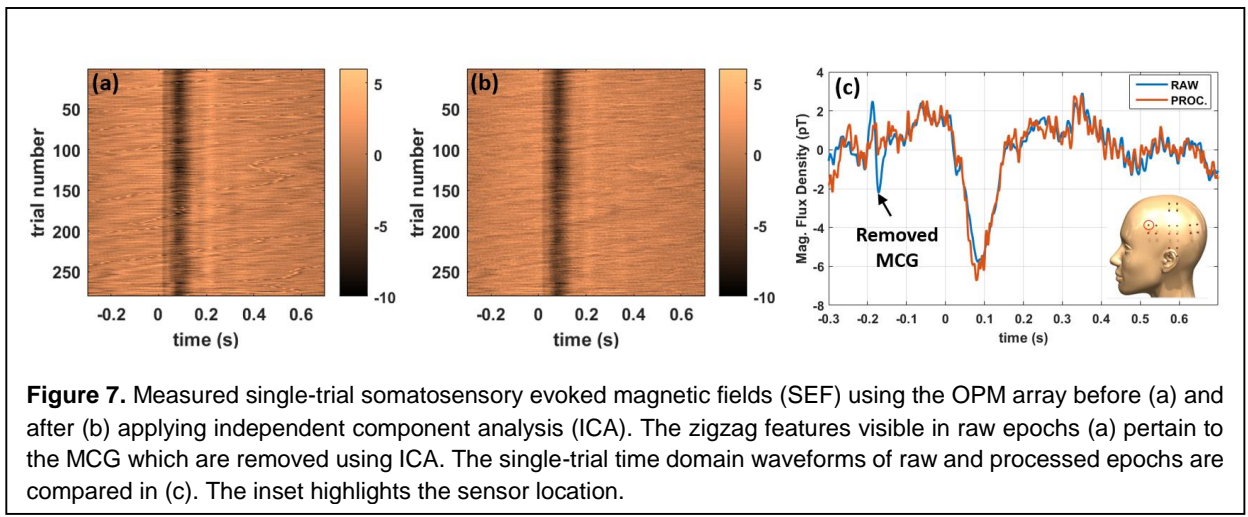

**Figure 7.** Measured single-trial somatosensory evoked magnetic fields (SEF) using the OPM array before (a) and after (b) applying independent component analysis (ICA). The zigzag features visible in raw epochs (a) pertain to the MCG which are removed using ICA. The single-trial time domain waveforms of raw and processed epochs are compared in (c). The inset highlights the sensor location.

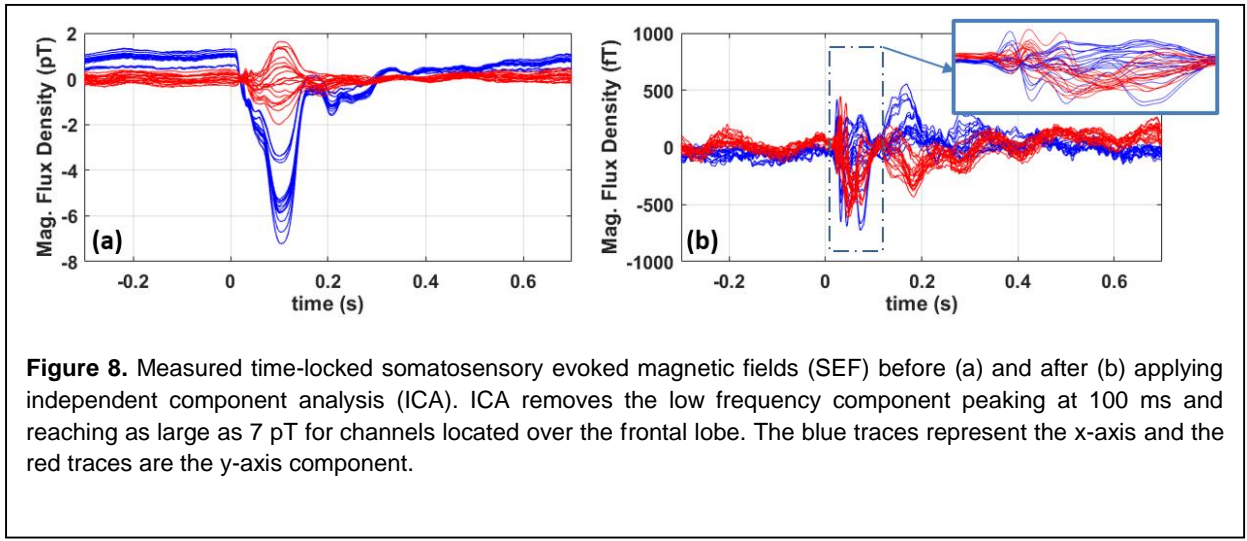

**Figure 8.** Measured time-locked somatosensory evoked magnetic fields (SEF) before (a) and after (b) applying independent component analysis (ICA). ICA removes the low frequency component peaking at 100 ms and reaching as large as 7 pT for channels located over the frontal lobe. The blue traces represent the x-axis and the red traces are the y-axis component.

### b. Somatosensory Evoked Magnetic Field (SEF)

The robustness of the measured SEF data from the OPM array is shown in figure 7. In this figure the individual epochs of the SEF data on a single channel are depicted over time for all the available trials before (a) and after (b) applying independent component analysis to remove only the MCG signal. Before ICA, in figure 7(a), the MCG signals are observed as zigzag features, and they are removed by the ICA as can be seen in figure 7(b). This distinction can be observed in the time domain with a single trial as shown in figure 7(c); the indicated MCG signal on the raw data, is no longer present on the processed data. In the OPM data a large 100 ms component, which was experimentally determined to be stemming from the left somatosensory cortex S2 in References [2, 33], was observed in all three subjects (see figure 8a) with a magnitude of greater than 5 pT. ICA is a necessary step for removing the large low-frequency event at 100 ms. In SEF data analysis, we look for the N20m and P30m components which have smaller amplitudes compared to the component at 100 ms to compare to the SQUID data. It should be noted that the ratio of the 100 ms component's amplitude to that of the N20m is an order of magnitude larger in the data recorded by the OPM array compared to that of the SQUID. ICA is capable of eliminating this large component, reaching 7 pT at around 100 ms, as shown in figure 8. For the rest of the SEF data analysis presented in this section, the S2 component and the MCG signal are removed using ICA.

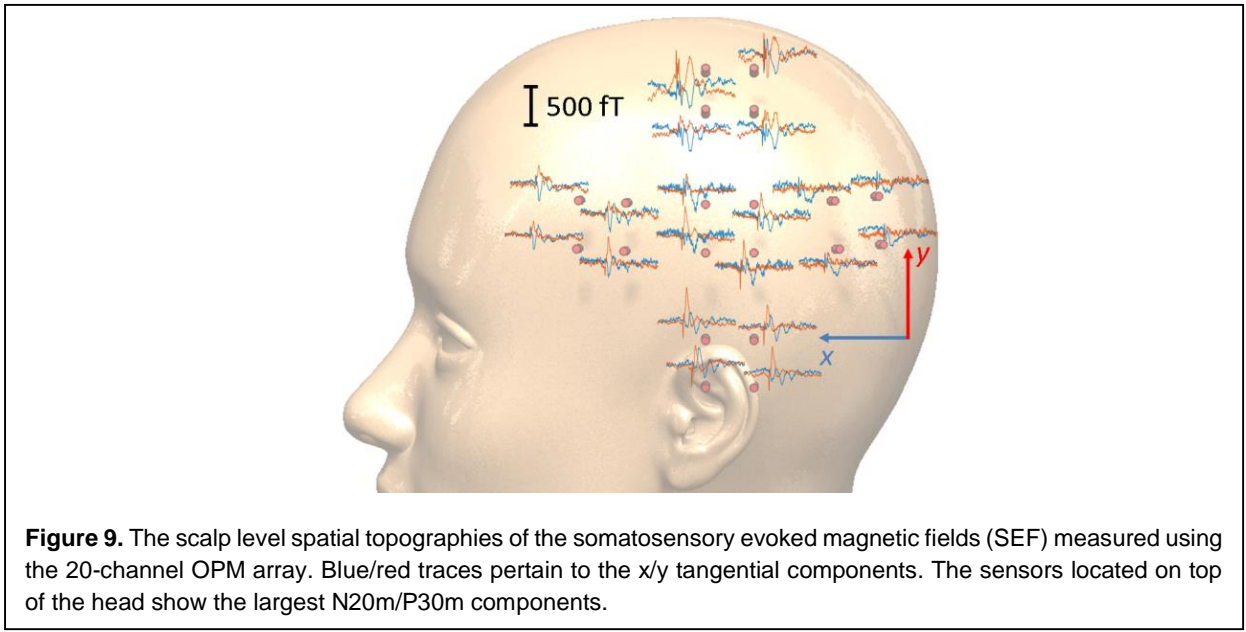

**Figure 9.** The scalp level spatial topographies of the somatosensory evoked magnetic fields (SEF) measured using the 20-channel OPM array. Blue/red traces pertain to the x/y tangential components. The sensors located on top of the head show the largest N20m/P30m components.

The average of the time-locked epochs to the peripheral stimulator's trigger are depicted in figure 9 for one subject. In this figure the ICA is used to remove both the MCG and the S2 components. Both x-axis and y-axis N20m and P30m components of SEF signals are largest for the sensor located above the somatosensory cortex S1, i.e. the sensor closer to the top of the head.

Figure 10(a) shows the averaged waveform of the SEF response for three subjects for both x-axis and y-axis modulation directions. The N20m and P30m peaks are indicated on figure 10(a). Figure 10(b) shows the fieldmaps of the SEF for the three subjects at the N20m peak; for all the three subjects, the largest amplitude response is located close to the top of the head. The comparison between the averaged waveforms of OPM and SQUID arrays is shown in figure 11. In figure 11, the SQUID channels are gradiometers; hence, in this figure, the SQUID signal is scaled arbitrarily to match the OPM signal. There is a correlation of 85 % and 75 % for the x and y tangential components respectively. Both OPM and SQUID sensor reveal similar time-frequency spectra as shown in figure 11 c-d; it is worth noticing that the beta suppression [34, 35] is observed with both SQUID and OPM array.

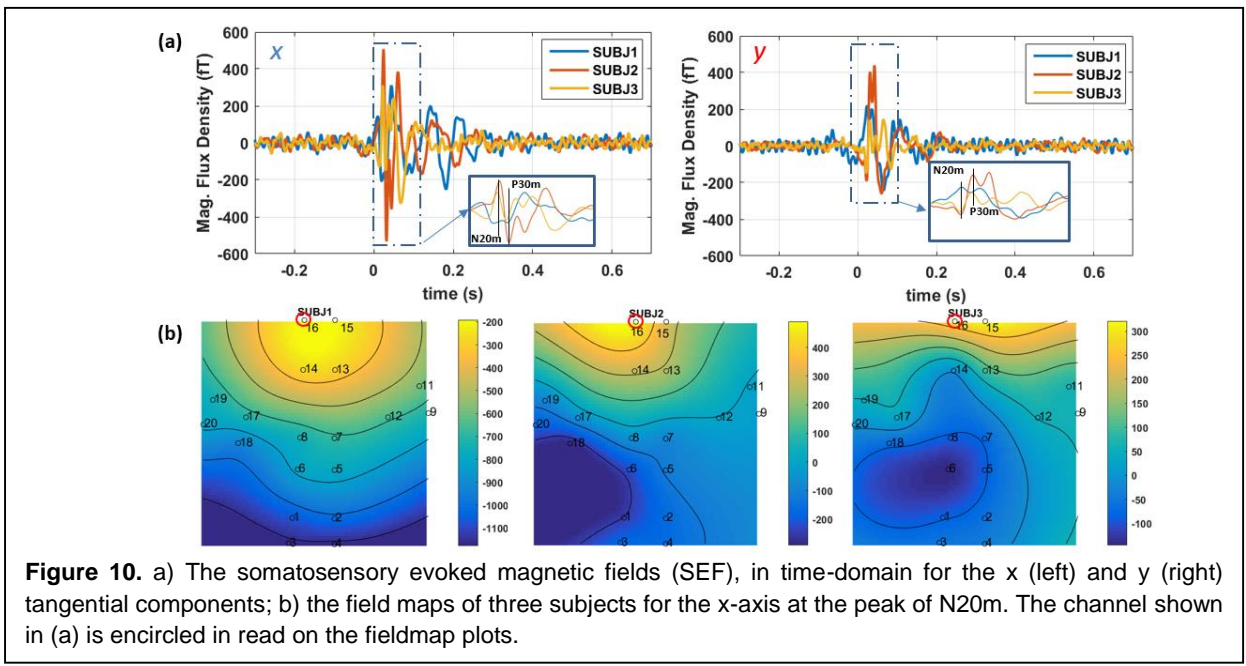

**Figure 10.** a) The somatosensory evoked magnetic fields (SEF), in time-domain for the x (left) and y (right) tangential components; b) the field maps of three subjects for the x-axis at the peak of N20m. The channel shown in (a) is encircled in read on the fieldmap plots.

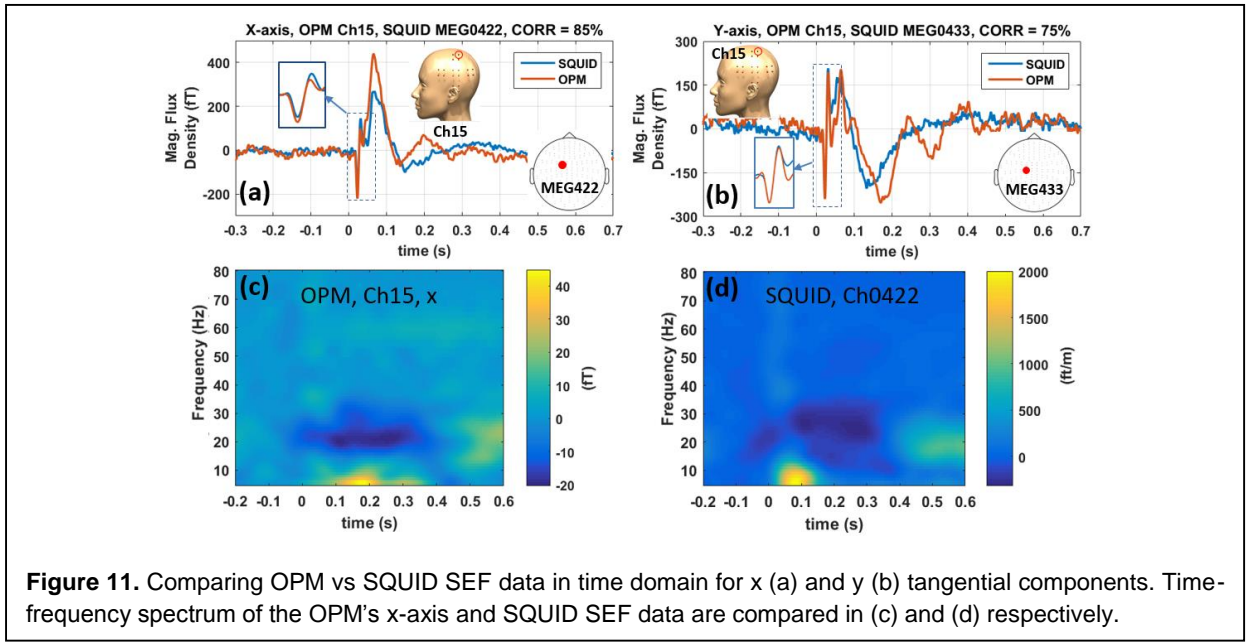

**Figure 11.** Comparing OPM vs SQUID SEF data in time domain for x (a) and y (b) tangential components. Time-frequency spectrum of the OPM's x-axis and SQUID SEF data are compared in (c) and (d) respectively.

## IV. Discussion and Conclusion

We have successfully developed a 20-channel OPM-based MEG system. We achieved high quality recording of AEF and SEF data with identification of standard peaks in the averaged evoked waveforms, while showing semi-quantitative similarity between the OPM- and SQUID-based recordings. It is important to note the large amplitude (7 pT) 100 ms component in the SEF. Other reports of SEF measurements with OPM show an amplitude of ~2 pT [8, 20]. Perhaps the large amplitude fields are only detectable with tangential measurement of the magnetic field. The large amplitude response and the low noise of our magnetic shield makes single trail observation of the signal possible with minimal data processing (only band-pass filtering). This provides support for the considerable advantage obtained through moving the sensors closer to the head. Through noise measurements using our OPMs, we can compare the noise performance of the three-layer MSR at the Mind Research Network [36] to that of our person-sized shield. We find the person-sized shield has three times less average noise from 5 to 100 Hz, 15 times less noise at 60 Hz, and a 100 times smaller residual field. The field maps generated by the OPM array's channels, for all the three subjects in both AEF and SEF experiments, show promising results for magnetic source localization through a clear variation in field strength across distance. Demonstrating source localization with our array is the subject of current research, and we believe that consistent source localization is readily achievable. Finally, our results indicate the possibility of creating larger arrays of OPMs for MEG and moving toward a conformable and flexible array of OPM sensors with full-head coverage.

**Acknowledgements**

The authors thank Kim Paulson for help in collecting the SQUID MEG data and Jeff Hunker for building the 18-coil system for the OPM MEG system. Sandia National Laboratories is a multi-mission laboratory managed and operated by Sandia Corporation, a wholly owned subsidiary of Lockheed Martin Corporation, for the U.S. Department of Energy's National Nuclear Security Administration under contract DE-AC04-94AL85000. Research reported in this publication was supported by the National Institute of Biomedical Imaging and Bioengineering of the National Institutes of Health under award number R01EB013302. The content is solely the responsibility of the authors and does not necessarily represent the official views of the National Institutes of Health.

**References**

1. Cohen, D., *MAGNETOENCEPHALOGRAPHY - EVIDENCE OF MAGNETIC FIELDS PRODUCED BY ALPHA-RHYTHM CURRENTS.* Science, 1968. **161**(3843): p. 784-&.
2. Hämäläinen, M., et al., *Magnetoencephalography--theory, instrumentation, and applications to noninvasive studies of the working human brain.* Rev. Mod. Phys., 1993. **65**(2): p. 413-497.
3. Leahy, R.M., et al., *A study of dipole localization accuracy for MEG and EEG using a human skull phantom.* Electroencephalography and Clinical Neurophysiology, 1998. **107**(2): p. 159-173.
4. Iivanainen, J., M. Stenroos, and L. Parkkonen, *Measuring MEG closer to the brain: Performance of on-scalp sensor arrays.* NeuroImage, 2017. **147**: p. 542-553.
5. Boto, E., et al., *On the Potential of a New Generation of Magnetometers for MEG: A Beamformer Simulation Study.* Plos One, 2016. **11**(8).
6. Xia, H., et al., *Magnetoencephalography with an atomic magnetometer.* Applied Physics Letters, 2006. **89**(21): p. 211104.
7. Kim, K., et al., *Multi-channel atomic magnetometer for magnetoencephalography: A configuration study.* NeuroImage, 2014. **89**: p. 143-151.
8. Sander, T.H., et al., *Magnetoencephalography with a chip-scale atomic magnetometer.* Biomed. Opt. Express, 2012. **3**(5): p. 981-990.
9. Johnson, C., P.D.D. Schwindt, and M. Weisend, *Magnetoencephalography with a two-color pump-probe, fiber-coupled atomic magnetometer.* Applied Physics Letters, 2010. **97**(24): p. 243703-3.
10. Shah, V.K. and R.T. Wakai, *A Compact, High Performance Atomic Magnetometer for Biomedical Applications.* Physics in Medicine and Biology, 2013. **58**(22): p. 8153-8161.
11. Kameda, K., et al., *Human magnetoencephalogram measurements using newly developed compact module of high-sensitivity atomic magnetometer.* Japanese Journal of Applied Physics, 2015. **54**(2).
12. Shah, V. and M.V. Romalis, *Spin-exchange relaxation-free magnetometry using elliptically polarized light.* Phys. Rev. A, 2009. **80**(1): p. 013416.
13. Higbie, J.M., E. Corsini, and D. Budker, *Robust, high-speed, all-optical atomic magnetometer.* Review of Scientific Instruments, 2006. **77**(11): p. 113106.
14. Mhaskar, R., S. Knappe, and J. Kitching, *A low-power, high-sensitivity micromachined optical magnetometer.* Applied Physics Letters, 2012. **101**(24).
15. Preusser, J., et al. *A microfabricated photonic magnetometer*. in *2009 IEEE International Frequency Control Symposium Joint with the 22nd European Frequency and Time forum*. 2009.
16. Schwindt, P.D.D., et al., *Chip-scale atomic magnetometer with improved sensitivity by use of the M-x technique.* Applied Physics Letters, 2007. **90**(8).
17. Dammers, J., et al., *Source localization of brain activity using helium-free interferometer.* Applied Physics Letters, 2014. **104**(21): p. 4.

18. Oisjoen, F., et al., *High-T-c superconducting quantum interference device recordings of spontaneous brain activity: Towards high-T-c magnetoencephalography.* Applied Physics Letters, 2012. **100**(13): p. 4.
19. Alem, O., et al., *Magnetic field imaging with microfabricated optically-pumped magnetometers.* Optics Express, 2017. **25**(7): p. 7849-7858.
20. Boto, E., et al., *A new generation of magnetoencephalography: Room temperature measurements using optically-pumped magnetometers.* NeuroImage, 2017. **149**: p. 404-414.
21. Colombo, A.P., et al., *Four-channel optically pumped atomic magnetometer for magnetoencephalography.* Optics Express, 2016. **24**(14): p. 15403-15416.
22. Kominis, I.K., et al., *A subfemtotesla multichannel atomic magnetometer.* Nature, 2003. **422**: p. 596-599.
23. Sullivan, G.W., et al., *A magnetic shielded room designed for magnetoencephalography.* Review of Scientific Instruments, 1989. **60**(4): p. 765-770.
24. Mager, A.J., *MAGNETIC SHIELDS.* Ieee Transactions on Magnetics, 1970. **MAG6**(1): p. 67-&.
25. Scofield, J.H., *Frequency-domain description of a lock-in amplifier.* American Journal of Physics, 1994. **62**(2): p. 129-133.
26. Okada, Y., et al., *BabyMEG: A whole-head pediatric magnetoencephalography system for human brain development research.* Rev Sci Instrum, 2016. **87**(9): p. 094301.
27. Cohentannoudji, C. and Dupontro.J, *EXPERIMENTAL STUDY OF ZEEMAN LIGHT SHIFTS IN WEAK MAGNETIC-FIELDS.* Physical Review a-General Physics, 1972. **5**(2): p. 968-+.
28. Oostenveld, R., et al., *FieldTrip: Open Source Software for Advanced Analysis of MEG, EEG, and Invasive Electrophysiological Data.* Computational Intelligence and Neuroscience, 2011. **2011**: p. 9.
29. Escudero, J., et al., *Artifact removal in magnetoencephalogram background activity with independent component analysis.* Ieee Transactions on Biomedical Engineering, 2007. **54**(11): p. 1965-1973.
30. Vigario, R., et al., *Independent component approach to the analysis of EEG and MEG recordings.* Ieee Transactions on Biomedical Engineering, 2000. **47**(5): p. 589-593.
31. Yamada, T., et al., *Asymmetrical enhancement of middle-latency auditory evoked fields with aging.* Neuroscience Letters, 2003. **337**(1): p. 21-24.
32. Virtanen, J., et al., *Replicability of MEG and EEG measures of the auditory N1/N1m-response.* Evoked Potentials-Electroencephalography and Clinical Neurophysiology, 1998. **108**(3): p. 291-298.
33. Kakigi, R., *SOMATOSENSORY-EVOKED MAGNETIC-FIELDS FOLLOWING MEDIAN NERVE-STIMULATION.* Neuroscience Research, 1994. **20**(2): p. 165-174.
34. Tiihonen, J., M. Kajola, and R. Hari, *MAGNETIC MU RHYTHM IN MAN.* Neuroscience, 1989. **32**(3): p. 793-800.
35. Pfurtscheller, G. and F.H.L. da Silva, *Event-related EEG/MEG synchronization and desynchronization: basic principles.* Clinical Neurophysiology, 1999. **110**(11): p. 1842-1857.
36. Johnson, C.N., P.D.D. Schwindt, and M. Weisend, *Multi-sensor magnetoencephalography with atomic magnetometers.* Physics in Medicine and Biology, 2013. **58**(17): p. 6065-6077.